# Mathematical Principles in Software Quality Engineering

Dr.Manoranjan Kumar Singh
PG Department of Mathematics
Magadh University
Bodhagaya - 823001, India
drmksingh_gaya@yahoo.com

Rakesh.L
Department of Computer-Science
S.C.T. Institute of Technology
Bangalore - 560075, India
rakeshsct@yahoo.co.in

*Abstract*— **Mathematics has many useful properties for developing of complex software systems. One is that it can exactly describe a physical situation of the object or outcome of an action. Mathematics support abstraction and this is an excellent medium for modeling, since it is an exact medium there is a little possibility of ambiguity. This paper demonstrates that mathematics provides a high level of validation when it is used as a software medium. It also outlines distinguishing characteristics of structural testing which is based on the source code of the program tested. Structural testing methods are very amenable to rigorous definition, mathematical analysis and precise measurement. Finally, it also discusses functional and structural testing debate to have a sense of complete testing. Any program can be considered to be a function in the sense that program input forms its domain and program outputs form its range. In general discrete mathematics is more applicable to functional testing, while graph theory pertains more to structural testing.**

I. INTRODUCTION

Software testing is one element of broader topic that is often referred to as verification and validation. Verification refers to the set of activities that ensure that software correctly implements a specific function. Verification methods ensure the system complies with an organization standards and processes, relying on review or non executable methods. Validation physically ensures that the system operates according to plan by executing the system functions through a series of tests that can be observed and evaluated. The advantage of mathematics is that it allows rigorous analysis and avoids an overreliance on intuition. Mathematics provides precise unambiguous statements and the mathematical proof of a theorem provides a high degree of confidence in its correctness. The emphasis is on mathematics that can be applied rather than mathematics for its own sake. The engineering approach aims to show how mathematics can be used to solve practical problems. The engineer applies mathematics and models to the design of the product and the analysis of the design is a mathematical activity. The use of mathematics will enable the software engineer to produce high quality products that are safe to use. The mathematics required by engineers include set theory, relations, functions, logics related to mathematics, tabular expressions, matrix theory, propositional logic, graph theory, finite state automata, calculus and probability theory. In general discrete mathematics is more applicable to functional testing, while graph theory pertains to structural testing. More than any other software life cycle activity testing lends itself to mathematical description and analysis. Testing of software is a means of measuring or assessing the software to determine its quality. Testing does provide the last bastion from which quality can be assessed and, more programmatically, errors can be uncovered [1]. But testing should not be viewed as a safety net. As experts say, "you can't test in quality. If it's not there before you begin testing, it won't be there when you're finished testing". Quality is incorporated into software throughout the process of software engineering. Miller relates software testing to quality assurance by stating that "the underlying motivation of program testing is to affirm software quality with methods that can be economically and effectively applied to both large scale and small scale systems" [8]. Verification and validation techniques can be applied to every element of the computerized system. Verification and Validation encompasses a wide array of software quality assurance activity that include formal technical reviews, quality and configuration audits, performance monitoring, simulation, feasibility study, documentation review, database review, algorithm analysis, development testing, usability testing, qualification testing and installation testing. The two broad categories of testing, functional testing and structural testing. Functional testing is sometimes called black box testing because no knowledge of the internal logic of the system is used to develop test cases. Structural testing is sometimes called white box testing because knowledge of the internal logic of the system is used to develop hypothetical test cases. Structural test uses verification predominantly. The properties that the test set is to reflect are classified according to whether they are derived from a description of the program function or from the program internal logic. Black box and white box testing exists from the definition. The paper is organized into different sections primarily focusing on the importance of discrete mathematics in verification and validation activity, testing life cycle, discrete math and its implication in identifying and analyzing test cases and making useful progression through results and conclusion.

## II. Discrete Math for testing

### A. Set Theory

The essence of software testing is to determine a set of test cases for the item to be tested. Set is a collection of well defined objects that contains no duplicates. For example, the set of natural numbers N is an infinite set consisting of the numbers 1, 2... and so on. Most sets encountered in computer science are finite as computers can only deal with finite entities. Set theory is a fundamental building of mathematics, Venn diagrams are often employed to give pictorial representation of a set and the various set operations. The two important variation of set are naive versus axiomatic set theory. In naive set theory, set is recognised as a primitive term, much like point and line which are primitive concept in geometry. Some of the synonyms for set are collection, group, bunch or a whole, for example instance we might wish to refer to the set of months that have exactly 30 days. In set notation it is represented as,

$$M1 = \{April, June, September, November\} \quad (1)$$

We read this notation as "M1 is the set whose elements are the Months April, June, September, and November". Sets are commonly pictured by Venn diagrams. In Venn diagrams, a set is depicted as a circle and points in the interior of the circle corresponds to elements of the set. Then, we might draw our set M1 of 30 day months as in Fig. 1.

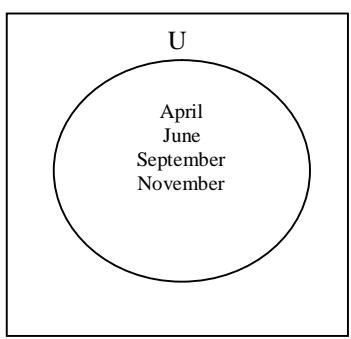

Fig. 1 Venn diagram for the set of 30 day Month

Venn diagrams communicate various set relationships in an intuitive way. Testing is fundamentally concerned with behaviour, and behaviour is orthogonal to the structural view common to software engineers. A quick differentiation is that structural view focuses on what it is and the behavioural view considers what it does. In this section we developed a simple Venn diagram that clarifies several nagging questions and is a useful mathematical tool for testing for graphical analysis.

### B. Graph Theory

Graph theory is a branch of topology that is sometimes as rubber sheet geometry. Curious, because rubber sheet parts of topology have little to do with graph theory. Furthermore graphs in the graph theory do not involve axes, scales, points, and curves as we might expect. Whatever the origin of the term, graph theory is probably the most useful part of mathematics for computer science. A graph also known as linear graph is an abstract mathematical structure defined from two, a set of nodes and set of edges that set form connections between nodes. A computer network is a fine example of a graph. A graph $G = (V, E)$ is composed of a finite (and non empty set) V of nodes and a set E of unordered pairs of nodes.

$$V = \{n_1, n_2, ..., n_m\},$$
$$\text{and}$$
$$E = \{e_1, e_2, ..., e_p\} \quad (2)$$

Where each edge $e_k = \{n_i, n_j\}$ for some nodes $n_i, n_j \in V$. From set theory the set $\{n_i, n_j\}$ is an unordered pair, which we sometimes write as $\{n_i, n_j\}$. Nodes are sometimes called vertices and edges are sometimes called arcs and we sometimes call nodes the end points of an arc. The common visual form of a graph shows nodes as circles and edges as lines connecting pair of nodes. To define a particular graph, we must first define a set of nodes and then define a set of edges between pairs of nodes. We usually think of nodes as program statements and we have various kinds of edges, representing, for instance, flow of control or define /use relations. The important property of graph which has deep implications for testing is cyclomatic complexity. The cyclomatic number of a graph is given by

$$V(G) = e - n + p, \quad \text{where}$$

E is the number of edges in G, N is the number of nodes in G, P is the number of components in G, V (G) is the number distinct regions in a graph.

One formulation of structural testing postulates the notion of basis paths in a program and shows that the cyclomatic number of a program graph is the number of these basis elements. There are four special graphs that are used for software verification. The first of these, the program graph, used primarily at the unit testing level. The other three, finite state machines, state charts, and petri nets are best used to describe system level behaviour, although they can be used at lower levels of testing. Program graph which is quite popular in testing can be defined as given a program written in an imperative programming language, its program graph is a directed graph in which, Nodes are programme statements, and edges represent flow of control, there is an edge from node i to node j if the statement corresponding to node j can be executed immediately after the statement corresponding to node i. In the graph in Figure 2 the node and edge set are:

$$V = \{n_1, n_2, n_3, n_4, n_5, n_6, n_7\}$$

$$E = \{e_1, e_2, e_3, e_4, e_5, e_6, e_7\}$$

$$= \{(n_1, n_2), (n_1, n_2), (n_1, n_4), (n_3, n_4), (n_2, n_5), (n_4, n_6)\} \quad (3)$$

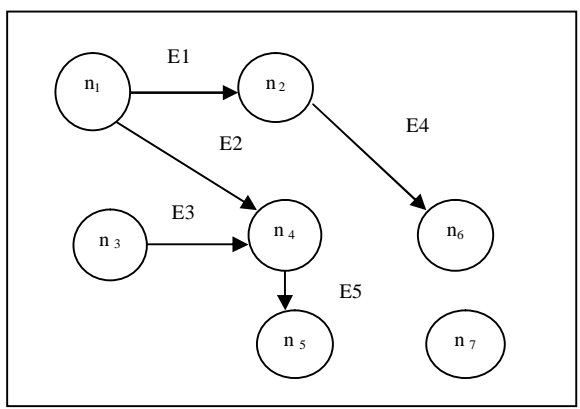

Fig.2 A graph with seven nodes and five edges

The directed graph formulation of program enables a very precise description of testing aspects of program. For one thing, a very satisfying connection exists between this formulation and the precepts of structured programming. The basic structured programming constructs sequence, selection, repetition all have a clear, directed graph [6]. When these constructs are used in structured program, the corresponding graphs are either nested or concatenated. The single entrance and single exit criteria result in unique source and sink nodes in the program graph. In fact, the old non structured "spaghetti code" resulted in very complex program graphs. GOTO statement, for example introduces edges and when these are used to branch into or out of loops, the resulting program graphs become even more complex.

One of the pioneering analysts of this is Thomas McCabe, who popularised the cyclomatic number of a graph as an indicator of program complexity (McCabe 1976). When program executes, the statement that execute comprise a path in the program graph. Loops and decision greatly increase the number of possible paths and therefore similarly increase the need for testing.

*C. Propositional Logic*

Propositional logic are concerned with propositional operators which may be applied to one or more propositions giving new propositions. A propositional variable is used to stand for a proposition, let P stand for the proposition '2 + 2 = 4' which is true and propositional variable can take either true or false. Each propositional variable has two possible values, and a formula with *n*-propositional variables has $2^n$ values associated with the propositional variables. The set of values associated with the *n* variables may be used to derive truth table with $2^n$ rows and *n + 1* columns. Each row gives each of $2^n$ values that the *n* variables may take and the column *n + 1* gives the result of the logical expression for that set of values of the propositional variable. Propositional logic allows further truth to be derived by logical reasoning or rules of inference. These rules enable new propositions to be deduced from a set of existing propositions provided that the rules of inference for the logic are followed. A valid argument or deduction is truth preserving i.e., if the set of propositions is true then the deduced proposition will also be true. It has also been applied to computer science and the term Boolean algebra is named after the English mathematician 'George Boole'. Boole was the first professor of mathematics at Queens College, Cork in mid $19^{th}$ century and he formalized the laws of propositional logic that are foundation for modern computers. Boolean algebra is widely used in programs for example, the Boolean condition in an *if then else* statement determines whether particular statement will be executed or not. Similarly, the Boolean condition in a while or for loop will determine if the statement in the body of the loop will be executed. Set theory, graph theory and propositional logic have chicken-and-egg relationship. A propositional logic expression can be viewed as composed of set of N elements, a truth function from N inputs to one output, a one-to-one correspondence between the N elements and the N inputs. For instance the expression:

$$((a \& c) | (b \& \sim e)) ==> ((c | m) <==> (m \& e))$$

Can be viewed as composed of a set of a 5 elements:
$$\{a, b, c, e, m\}$$

A truth function $f(x_1, x_2, x_3, x_4, x_5)$ :
$$f(x_1, x_2, x_3, x_4, x_5) = ((x_1 \& x_2) | (x_3 \& \sim x_4)) ==> ((x_2 | x_5) <==> (x_5 \& x_4)) \quad (4)$$

And one-to-one correspondence can be written as :

$$\{(a, x_1), (b, x_3), (c, x_3), (e, x_4), (m, x_5)\} \quad (5)$$

The resulting expression can be represented graphically.

III. A TESTING LIFE CYCLE

Much of the testing literature is mired in confusing terminology, probably because testing technology evolved over decades. The following terminology is taken from standards developed from the institute of Electronics and Electrical Engineers (IEEE) Computer Society [3].

**Error**: People make errors. A good synonym is mistake. When people make mistakes during coding, we call these mistakes bugs. Errors tend to propagate, a requirements error may be magnified during design and amplified still more during coding.

**Fault:** A fault is the result of an error. It is more precise to say that fault is the representation of an error. Defect is a good synonym for fault, as is bug. Faults can be elusive.

**Failure:** A failure occurs when a fault executes. Two subtleties arise here, one is that failures only occur in an executable representation, which is usually taken to be source code or more precisely loaded object code. The second subtlety is this definition relates failures only to faults of commission and not faults of omission. Also, there are faults that never happen to execute, or perhaps do not execute for a long time. The Michelangelo virus is an example of such a fault. It does not execute until Michelangelo's birthday, March 6th. Reviews prevent many failures by finding fault.

**Incident:** When failure occurs, it may or may not be readily apparent to user. An incident is the symptom associated with a failure that alerts the user to the occurrence of a failure

**Test:** Testing is obviously concerned with errors, faults, failures and incidents. Test is the act of exercising software with test cases. A test has two distinct goals to find failures or to demonstrate correct execution.

**Test case:** Test case has a identity and is associated with a program behaviour. A test case also has a set of inputs and a list of expected outputs.

Software development involves ambiguity, assumptions and flawed human communication. Each change made to a piece of software, each new piece of functionality, each attempt to fix a defect, introduces the possibility of error [7]. With each error, the risk that the software will not fulfill its intended purpose increases. Testing reduces that risk. We can use quality assurance processes to attempt to prevent defects from entering software but the only thing we can do to reduce the number of errors already present is to test it. By following a cycle of testing and rectification we can identify issues and resolve them. The life cycle model of testing in depicted in the Figure 3. Notice that, in the development phases, three opportunities arise for errors to be made resulting in faults that propagate through the remainder of the development process. The activity in the life cycle can be summarized as, the first three phases are putting bugs in, the testing phase is finding bugs and the last three phases are getting bugs out. The fault resolution step is another opportunity for errors. When fix causes formerly correct software to misbehave, the fix is deficient. From this sequence of terms, we see that test cases occupy a central position in testing.

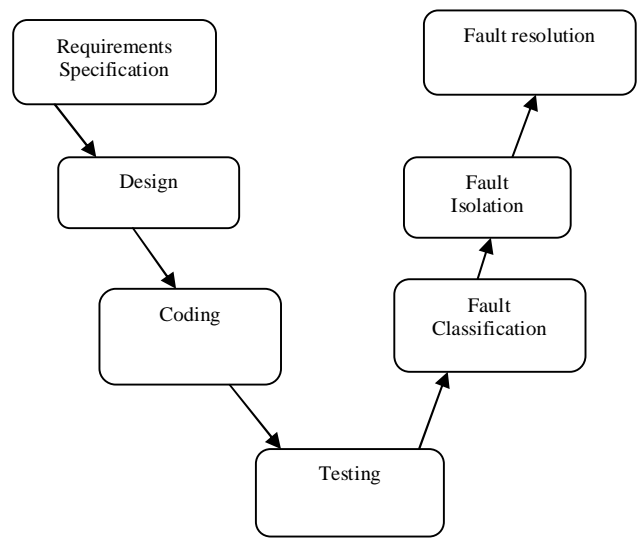

Fig. 3 Testing Life cycle

There are different levels of testing, which reflects the levels of abstraction found in waterfall model of software development life cycle. Waterfall model can be thought of as a linear sequence of events. We start at A, we do B and then go to C and eventually end up at Z. This is extremely simplistic but it does allow you to visualise the series of events in the simplest way. Unit testing, Integration testing and System testing are different levels of testing. A practical relationship exists between levels of testing versus functional and structural testing [4]. Structural testing is more appropriate at unit level, while functional testing is more appropriate at system level. This is generally true, but is also likely consequence of the base information produced during the requirements specification, preliminary design, and detailed design phases. The constructs defined for structural testing make the most sense at the unit level, and similar constructs are only now becoming available for the integration and system levels of testing.

IV. FINDINGS AND OBSERVATIONS

Consider a universe of program behaviours, given a program and its specification. Consider the set S of specified behaviours, and the set P of programmed behaviours. Figure 4 shows the relation the relationship among our universe of discourse as well as the specified and programmed behaviours.

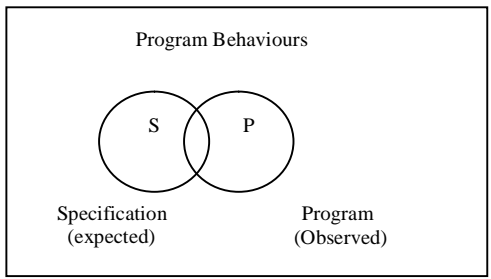

Fig. 4 Specified and implemented program behaviour

Of all the possible program behaviours, the specified ones are in the circle labelled S and all those behaviours actually programmed are in P. With this diagram we can see more clearly the problem that confront a tester. What if certain specified behaviour have not been programmed, in our earlier terminology, these are faults of omission. Similarly, what if certain programmed or implemented behaviour have not been specified. This corresponds to faults of commission and to errors that occurred after the specification was complete. The intersection of S and P the foot ball shaped region is the correct portion that is, the behaviours that are both specified and implemented. A very good view of testing is that it is the determination of the extent of program behaviour that is both specified and implemented. As an aside, note that correctness only has meaning with respect to specification and an implementation. It is a relative term, not an absolute. The representation in the Figure 5 gives effectiveness for identifying test cases. Notice that a slight discrepancy with our universe of discourse and the set of program behaviours. Now, consider the relationships among the sets S, P, and T. There may be specified behaviours that are not tested (region 2 and 5), specified behaviours that are tested (regions 1 and 4), and test cases that correspond to unspecified behaviours (region 3 and 7). Similarly, there may be programmed behaviours that are not tested (regions 2 and 6), programmed behaviours that are tested (region 1 and 3), and test cases that correspond to unprogrammed behaviours (regions 4 and 7). Each of these regions is important. If specified behaviours exist for which no test cases are available testing is necessarily incomplete. If certain test cases correspond to unspecified behaviours, two possibilities arise either such a test case is unwarranted or the specification is deficient. We have already arrived at a point where we can see some possibilities for testing as a craft: what can a tester do to make the region where these sets all intersect (region 1) as large as possible. Another approach is to ask how the test cases in the set T are identified. The short answer is that test cases are identified by a testing method. This frame work gives us a way to compare the effectiveness of diverse testing methods.

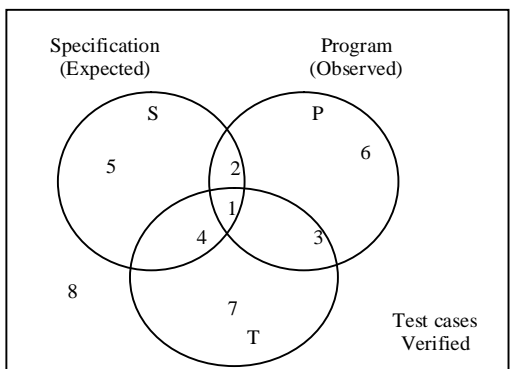

Fig. 5 Specified, implemented and tested behaviours

Venn diagrams are used in software development and testing life cycle, together with directed graph they are the basis of the state chart notations, which are among the most rigorous specification techniques supported by CASE technology. State charts are also the control notation chosen for the UML, Universal Modelling language from rational corporation and the Object management group. David Harel has two goals when he developed the state chart notation, he wanted to devise a visual notation that combined the ability of Venn diagrams to express hierarchy and the ability of directed graphs to express connectedness. David Harel uses the methodology neutral term "blob" to describe the basic building block of state chart. Blobs can contain other blobs in the same way that Venn diagrams show set containment [2]. Blobs can also be connected to other blobs with edges in the same way that the nodes in a directed graph are connected. We can interpret blobs as states, and edges as transitions. The full state chart system supports an elaborate language that defines how and when transition occurs. State charts are executable in much more elaborate way than ordinary finite state machines. Executing a state chart requires a notion similar to that of Petri net markings. The initial state of a state chart is indicated by an edge that has no source state. When states are nested with other states, the same indication is used to show the lower level initial state. Taken together these capabilities provide an elegant answer to the state explosion problem of ordinary finite state machines. The result is highly sophisticated and very precise notation for static analysis of software testing.

## V. RESULTS AND DISCUSSION

Functional testing is based on the view that any program can be considered to be a function that maps values from its input domain to values in its output range. Functions are a central notion to software development and testing. The whole functional decomposition paradigm, for example, implicitly uses the mathematical notion of a function. Informally function associates elements of a set. Any program can be thought of as a function that associates its output with its inputs. In the mathematical formulation of a function, the inputs are the domain and the outputs are the range of the function.

Given sets A and B, a function f is subset of A×B such that,

for $a_i, a_j \in A$, $b_i, b_j \in B$, and

$$f(a_i) = b_i,\ f(a_j) = b_j,\ b_i \neq b_j \Rightarrow a_i \neq a_j \qquad (6)$$

Formal definitions like this one are notoriously terse, so let us take a closer look. The inputs to the function f are elements of the set A, and the outputs of f are elements of B. What the definition says is that the function f is well behaved in the sense that an element in A is never associated with more than one element of B. In the above definition just given, the set A is the domain of the function f, and the set B is the range. Because input and output have a natural order, it is an easy step to say that a function f is really a set of ordered pairs in which the first element is in the domain. The notation for the function can be written as f: A → B, function are further described by particulars of the mapping. In the definition below, we start with a function f: A → B, we define the set:

$$f(A) = \{\ b_i \in B\ :\ b_i = f(a_i)\ \text{for some}\ a_i \in A\ \} \qquad (7)$$

This set is called the image of A under f.

Function, domain and range are commonly used in engineering. Black box testing is one in which the content (implementation) is not known, and the function of the black box is understood completely in terms of its inputs and outputs. Many times we operate very effectively with black box knowledge in fact this idea is central to the object orientation. With the functional approach to test case identification, the only information used is the specification of the software. Functional test cases have two distinct advantages, they are independent of how the software is implemented, so if the implementation changes, the test case are still useful and test case development can occur in parallel with the implementation thereby reducing overall project development interval [5]. On the negative side functional test cases frequently suffer from two problems, significant redundancies may exist among test cases, compounded by the possibility of gaps of untested software. Figure. 6 shows the mathematical analysis and representation of test cases identified by two functional methods in a broader perspective.

Method A identifies larger set of test cases than does method B. Notice that, for both methods, the set of test cases is completely contained within the set of specified behaviour.

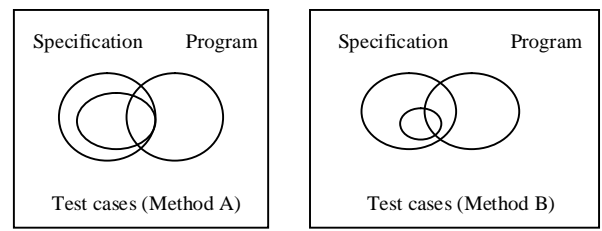

Fig. 6 Comparing Functional Test case methods (Mathematical View)

Because functional methods are based on the specified behaviour, it is hard to imagine these methods identifying behaviours that are not specified.

Structural testing is the other fundamental approach to test case identification To contrast it with the functional testing, its even sometimes called white box or even clear box. The clear box metaphor is probably more appropriate, because essential difference is that the implementation is known and used to identify test cases. The ability to see inside the black box allows the tester to identify test cases based on how the function is actually implemented .Structural testing has been the subject of some fairly strong theory. To really understand the structural testing the concepts of linear graph theory is essential. Because of the strong theoretical basis, structural testing lends itself to the definition and use of test coverage metrics. Figure. 7 shows the result of test cases identified by two structural methods.

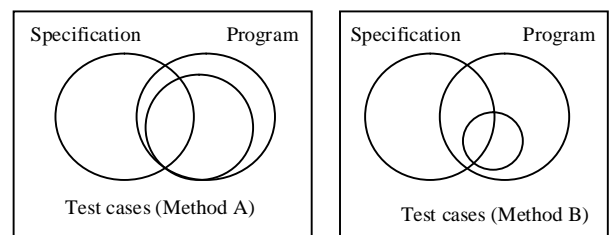

Fig. 7 Comparing structural Test case methods (Mathematical View)

As before, Method A identifies larger set of test cases as does Method B. We can notice that, for both methods, the set of test cases completely contained within the set of programmed behaviour. Because structural methods are based on program, it is hard to imagine these methods identifying behaviour that are not programmed. It is easy to imagine, however, that the set of structural test cases is relatively small with respect to the full set of programmed behaviours.

## V. CONCLUSION

In this paper authors used formal methods to solve practical problems of software quality engineering. Uncertainty is inherent and inevitable in software development process and products. Testing like any other development activity is human intensive and thus introduces uncertainties. The focus of set theory, graph theory and propositional logic which are fundamental principle of mathematics are applied to solve practical problems in developing a quality product. This proved to be well suited to a life cycle approach of software testing process that ensures, consistency, communicability, economy and an attractive modelling technique. Also, alternative methods are used to compare the fundamental nature of different testing techniques in identifying test cases with rigorous analysis and graphical representation is a highly useful tool. Finally, the rich body of classical mathematics applied to software engineering will definitely improve the development process by reducing the information loss and provide an effective alternative strategy in early stages of the development.

AUTHORS PROFILE

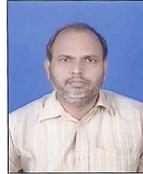

**Dr.Manoranjan Kumar Singh** received his Ph.D degree from Magadh University, India in 1986. This author is Young Scientist awardee from Indian Science Congress Association in 1989. A life member of Indian Mathematical society, Indian Science congress and Bihar Mathematical society. He is also the member of Society of Fuzzy Mathematics and Information Science, I.S.I. Kolkata. He was awarded best lecturer award in 2005. He is currently working as a Senior Reader in post graduation Department of Mathematics, Magadh University. He is having overall teaching experience of 26 years including professional colleges. His major research Interests are in Fuzzy logic, Expert systems, Artificial Intelligence and Computer-Engineering. He has completed successfully Research Project entitled, Some contribution to the theory of Fuzzy Algebraic Structure funded by University Grants Commission, Kolkata region, India. His ongoing work  includes Fuzzification of various Mathematical concepts to solve real world problems.

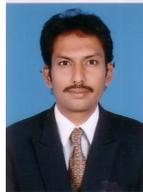

**L.Rakesh** received his M.Tech in Software Engineering from Sri Jayachamarajendra College of Engineering, Mysore, India in 2001 as an honour student. He is a member of International Association of Computer Science and Information Technology, Singapore. He is also a member of International Association of Engineers, Hong Kong. He is pursuing his Ph.D degree from Magadh University, India. Presently he is working as a Assistant professor and Head of the Department in Computer Science & Engineering, SCT Institute of Technology, Bangalore, India. He has presented and published research papers in various National, International conferences and Journals. His research interests are Formal Methods in Software Engineering, 3G-Wireless Communication, Mobile Computing,  Fuzzy Logic and Artificial agents.